\documentstyle[12pt,epsf,rotate]{article}

\begin{document}

\thispagestyle{empty}
\begin{flushright}
CERN-TH/98-261\\
BUTP-98/20\\
MPI-PhT/98-58\\
DESY 98-101\\
hep-ph/9808385\\
August 1998

\end{flushright}
\vspace*{0.8cm}
\centerline{\Large\bf Next-to-Leading Order QCD Corrections}
\vspace*{0.3cm}
\centerline{\Large\bf to the Lifetime Difference of $B_s$ Mesons}
\vspace*{1.5cm}
\centerline{{\sc M. Beneke},\ \ {\sc G. Buchalla}}
\smallskip
\centerline{\sl Theory Division, CERN, CH-1211 Geneva 23,
                Switzerland}
\bigskip
\centerline{{\sc C. Greub}} 
\smallskip
\centerline{\sl Institut f\"ur Theoretische Physik, Universit\"at Bern}
\centerline{\sl Sidlerstrasse 5, CH-3012 Berne, Switzerland}
\bigskip
\centerline{{\sc A. Lenz}}
\smallskip
\centerline{\sl Max-Planck-Institut f\"ur Physik -- 
                Werner-Heisenberg-Institut}
\centerline{\sl F\"ohringer Ring 6, D-80805 Munich, Germany}
\medskip
\centerline{and}
\medskip
\centerline{{\sc U. Nierste}}
\smallskip
\centerline{\sl DESY - Theory Group, Notkestra\ss e 85, 
                D-22607 Hamburg, Germany}
\vspace*{1cm}
\centerline{\bf Abstract}
\vspace*{0.3cm}
\noindent We compute the QCD corrections to the decay rate difference
in the $B_s$--$\bar B_s$ system, $\Delta\Gamma_{B_s}$, in the 
next-to-leading logarithmic approximation using the heavy quark 
expansion approach. Going beyond leading order in QCD is essential
to obtain a proper matching of the Wilson coefficients to the
matrix elements of local operators from lattice gauge theory. The lifetime 
difference is reduced considerably at next-to-leading order.
We find $(\Delta\Gamma/\Gamma)_{B_s}=(f_{B_s}/210\,{\rm MeV})^2
[0.006 B(m_b)+0.150 B_S(m_b)-0.063]$ in terms of the bag parameters
$B$, $B_S$ in the NDR scheme.
As a further application of our analysis we also derive the next-to-leading 
order result for the mixing-induced CP asymmetry in inclusive
$b\to u\bar ud$ decays, which measures $\sin 2\alpha$.

\vspace*{1cm}

\vfill

\newpage
\pagenumbering{arabic}

%\section{Introduction}

\noindent
{\bf 1. Introduction.} 
The width difference $(\Delta\Gamma/\Gamma)_{B_s}$ of the 
$B_s$ meson CP eigenstates \cite{LO} is expected to
be about $10-20\%$, among the largest rate differences in the $b$-hadron
sector \cite{BIG}, and might be measured in the near future.
A measurement of a sizeable $(\Delta\Gamma/\Gamma)_{B_s}$ would open up
the possibility of novel CP violation studies with $B_s$ mesons 
\cite{DUN1,DF}. In principle, a measured value
for $\Delta\Gamma_{B_s}$ could also give some information on the
mass difference $\Delta M_{B_s}$ \cite{BP}, if the theoretical prediction
for the ratio $(\Delta\Gamma/\Delta M)_{B_s}$ can be sufficiently
well controlled \cite{BBD1}.
Furthermore, as pointed out in \cite{GRO}, if non-standard-model sources
of CP violation are present in the $B_s$ system, $\Delta\Gamma_{B_s}$
can be smaller (but not larger) than expected in the standard model.
For this reason a lower bound on the standard model prediction is
of special interest.

The calculation of inclusive non-leptonic $b$-hadron decay observables,
such as $\Delta\Gamma_{B_s}$, uses the heavy quark expansion (HQE). The 
decay width difference is expanded in powers of $\Lambda_{\rm QCD}/m_b$, 
each term being multiplied by a series of radiative corrections in 
$\alpha_s(m_b)$. In the case of $(\Delta\Gamma/\Gamma)_{B_s}$, 
the leading contribution is parametrically of order 
$16\pi^2 (\Lambda_{\rm QCD}/m_b)^3$. In the framework of the HQE
the main ingredients for a reliable
prediction (less than $10\%$ uncertainty) are a) subleading corrections
in the $1/m_b$ expansion, b) the non-perturbative matrix elements of
local four-quark operators between $B$-meson states and c)
${\cal O}(\alpha_s)$ radiative corrections to the Wilson coefficients
of these operators. The first issue has been addressed in \cite{BBD1}.
The hadronic matrix elements can be studied using numerical simulations
in lattice QCD. In this letter, we present the next-to-leading 
order QCD radiative corrections to the Wilson coefficient functions for
$\Delta\Gamma_{B_s}$. In addition to removing another item from the 
above list and reducing certain renormalization scale
ambiguities of the leading order prediction, the inclusion
of ${\cal O}(\alpha_s)$ corrections is necessary for a satisfactory
matching of the Wilson coefficients to the matrix elements to be
obtained from lattice calculations.

Our results provide the first calculation of perturbative QCD effects
beyond the leading logarithmic approximation to spectator effects 
in the HQE for heavy hadron decays. The consideration of subleading
QCD radiative effects has implications of conceptual interest for
the construction of the HQE. Soft gluon emission from the 
spectator $s$ quark in the $B_s$ meson leads to power-like IR 
singularities in individual contributions, which would apparently 
impede the HQE construction, because they cannot be absorbed into 
matrix elements of local operators. It has already 
been explained in \cite{BU1}, how these severe IR divergences 
cancel in the sum over all cuts of a given diagram, so 
that the Wilson coefficients of four-quark operators relevant to 
lifetime differences, such as
$\Delta\Gamma_{B_s}$, are free of infrared singularities. This 
infrared cancellation is confirmed by the result of our explicit 
calculation.

Using the HQE to finite order in
$\Lambda_{QCD}/m_b$ rests on the assumption of 
local quark-hadron duality. Little is known in QCD about the
actual numerical size of duality-violating effects. 
(See, e.g., \cite{BSUV} for a 
recent discussion of the issue.)  
Experimentally no violation of local quark-hadron duality
in inclusive observables of the $B$-meson sector has 
been established so far.
In \cite{ALE93} it has been shown that for $\Delta\Gamma_{B_s}$
local duality holds exactly in the simultaneous limits of small
velocity ($\Lambda_{QCD}\ll$ $m_b-2m_c\ll m_b$) and large number of
colours ($N_c\to\infty$). In this case
\begin{equation}\label{dglim}
\left(\frac{\Delta\Gamma}{\Gamma}\right)_{B_s}=
\frac{G^2_F m^3_b f^2_{B_s}}{4\pi}|V_{cs}V_{cb}|^2\,
\sqrt{2-4\frac{m_c}{m_b}}\, \tau_{B_s}\approx 0.18.
\end{equation}
It is interesting that the numerical value implied
by the limiting formula (\ref{dglim}) appears to be quite realistic and 
is in
fact consistent with the results of more complete 
analyses \cite{BBD1,ALE93}. The duality assumption 
can in principle be tested by a
confrontation of theoretical predictions, based on the HQE,
with experiment. This aspect is another major motivation for 
computing $\Delta\Gamma_{B_s}$ accurately.

It is clear from these remarks that a detailed theoretical analysis of
$\Delta\Gamma_{B_s}$, and particularly of ${\cal O}(\alpha_s)$ 
corrections, is very desirable, both for phenomenological and
for conceptual reasons. In this letter we shall concentrate on the
presentation of our results and a brief discussion of their main
aspects. Details and an extension of our analysis to other
$b$-hadron lifetime differences will be given in a forthcoming
publication \cite{BBGLN2}.

\vspace*{0.4cm} 

%\section{Formalism and Results}

\noindent
{\bf 2. Formalism and next-to-leading order results.} 
In the limit of CP conservation the mass eigenstates of the
$B_s$--$\bar B_s$ system are 
$|B_{H/L}\rangle=(|B_s\rangle\pm |\bar B_s\rangle)/\sqrt{2}$, using the
convention $CP|B_s\rangle=-|\bar B_s\rangle$. The width difference
between mass eigenstates is then given by
\begin{equation}
\label{dgdef}
\Delta \Gamma_{B_s}\equiv\Gamma_L-\Gamma_H=
-2\,\Gamma_{12}=-2\,\Gamma_{21},
\end{equation}
where $\Gamma_{ij}$ are the elements of the decay-width matrix,
$i,j=1,2$ 
($|1\rangle=|B_s\rangle$, $|2\rangle=|\bar B_s\rangle$).
In writing (\ref{dgdef}) we assumed standard CKM phase conventions
\cite{PDG}. For more information about the basic formulas see
for instance \cite{BBD1} (and references therein).

The decay width is related to the absorptive part of the forward 
scattering amplitude via the optical theorem \cite{BIG}. 
The off-diagonal element of the decay-width matrix may thus be 
written as
\begin{equation}\label{g21t}
\Gamma_{21}=\frac{1}{2M_{B_s}}
\langle\bar B_s|{\cal T}|B_s\rangle.
\end{equation}
The normalization of states is 
$\langle B_s|B_s\rangle=2EV$ (conventional relativistic normalization)
and the transition operator ${\cal T}$ is defined by \cite{BIG}
\begin{equation}\label{tdef}
{\cal T}=\mbox{Im}\,i\int d^4x\ T\,{\cal H}_{eff}(x){\cal H}_{eff}(0).
\end{equation}
Here ${\cal H}_{eff}$ is the low energy effective weak
Hamiltonian mediating bottom quark decay. The component that is
relevant for $\Gamma_{21}$ reads explicitly
\begin{equation}\label{hpeng}
{\cal H}_{eff}=\frac{G_F}{\sqrt{2}}V^*_{cb}V_{cs}
\left(\,\sum^6_{r=1} C_r Q_r + C_8 Q_8\right),
\end{equation}
with the operators
\begin{equation}\label{q1q2}
Q_1= (\bar b_ic_j)_{V-A}(\bar c_js_i)_{V-A}\qquad
Q_2= (\bar b_ic_i)_{V-A}(\bar c_js_j)_{V-A},
\end{equation}
\begin{equation}\label{q3q4}
Q_3= (\bar b_is_i)_{V-A}(\bar q_jq_j)_{V-A}\qquad
Q_4= (\bar b_is_j)_{V-A}(\bar q_jq_i)_{V-A},
\end{equation}
\begin{equation}\label{q5q6}
Q_5= (\bar b_is_i)_{V-A}(\bar q_jq_j)_{V+A}\qquad
Q_6= (\bar b_is_j)_{V-A}(\bar q_jq_i)_{V+A},
\end{equation}
\begin{equation}\label{q8}
Q_8= \frac{g}{8\pi^2}m_b\, 
\bar b_i\sigma^{\mu\nu}(1-\gamma_5)T^a_{ij} s_j\, G^a_{\mu\nu}.
\end{equation}
Here the $i,j$ are colour indices and a summation over 
$q=u$, $d$, $s$, $c$, $b$ is implied. 
$V\pm A$ refers to $\gamma^\mu(1\pm\gamma_5)$ and $S-P$ (which 
we need below) to $(1-\gamma_5)$.
$C_1,\ldots, C_6$ are the corresponding Wilson coefficient
functions, which are known at next-to-leading order. 
We have also included the chromomagnetic operator $Q_8$, contributing
to ${\cal T}$ at ${\cal O}(\alpha_s)$. 
(Note that for a negative $C_8$, as conventionally used in the literature,
the Feynman rule for the quark-gluon vertex
is $-ig\gamma_\mu T^a$.)  
A detailed review and explicit expressions may be found in \cite{BBL}.
Cabibbo suppressed channels have been neglected in (\ref{hpeng}).

\begin{figure}[t]
\begin{center}
\vspace{0cm}
\hspace*{0.0cm}
  \epsfysize=0.98\textwidth
%\centerline{\rotate[r]{\epsfxsize=0.6\textwidth \epsffile{figs_nlo3.ps}}}
%   \centerline{\epsffile{figs_nlo2.ps}}
% \epsffile{figs_nlo3.ps}
       \rotate[r]{\epsffile{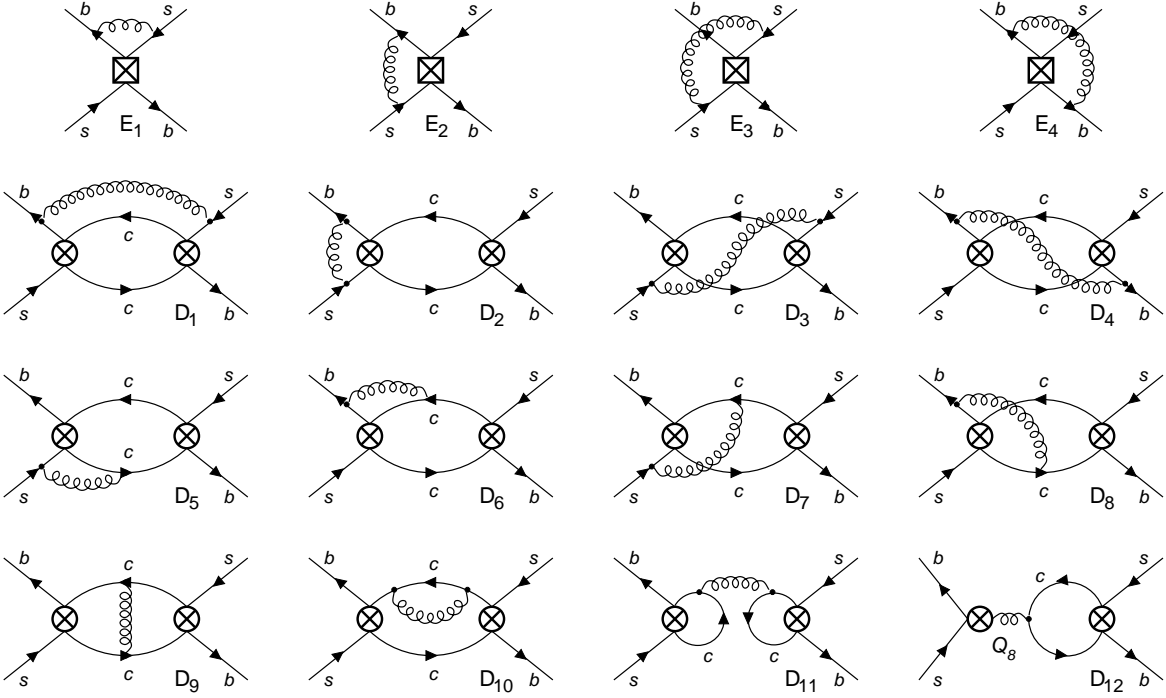}} 
\vspace*{-0.3cm}
\end{center}
\caption{${\cal O}(\alpha_s)$ corrections to the $B_s$--$\bar B_s$
  transition operator, $D_1$ -- $D_{12}$. Also shown are the 
  corrections to the matrix elements of local $\Delta B=2$ operators,
  $E_1$ -- $E_4$, required for a proper factorization of short-distance
  and long-distance contributions. Not displayed explicitly are
  $E_1'$, $E_2'$, $D_1'$, $D_2'$, $D_5'$, $D_6'$, $D_7'$, $D_8'$, $D_{10}'$
  and $D_{12}'$, which are obtained by rotating the corresponding
  diagrams by $180^\circ$.}
\protect\label{fignlo}
\end{figure}

Expanding the operator product (\ref{tdef}) for small $x\sim 1/m_b$, 
the transition operator ${\cal T}$ can be written, to leading order 
in the $1/m_b$ expansion, as
\begin{equation}\label{tfq}
{\cal T}=-\frac{G^2_F m^2_b}{12\pi}(V^*_{cb}V_{cs})^2
\, \left[ F(z) Q(\mu_2)+ F_S(z) Q_S(\mu_2) \right]
\end{equation}
with $z=m_c^2/m_b^2$ and the basis of $\Delta B=2$ operators
\begin{equation}\label{qqs}
Q = (\bar b_is_i)_{V-A}(\bar b_js_j)_{V-A},\qquad
Q_S= (\bar b_is_i)_{S-P}(\bar b_js_j)_{S-P} ~.
\end{equation}
In writing (\ref{tfq}) we have used Fierz identities and the equations 
of motion to eliminate the colour re-arranged operators
\begin{equation}\label{qqt}
\tilde Q=(\bar b_is_j)_{V-A}(\bar b_js_i)_{V-A},\qquad
\tilde Q_S= (\bar b_is_j)_{S-P}(\bar b_js_i)_{S-P}~,
\end{equation}
always working to leading order in $1/m_b$ (see below).
The Wilson coefficients $F$ and $F_S$ can be extracted
by computing the matrix elements between quark states of
${\cal T}$ in (\ref{tdef}) (Fig.~\ref{fignlo}), 
%%%%%%%%%%%%%%%%%%%%%%%%%%%%%%%%%%%%%%%%%%%%%%%%%%%%%%%%%%%%%%%%%%%
%%%%%%%%%%%%%%%%%%%%%%%%%%%%%%%%%%%%%%%%%%%%%%%%%%%%%%%%%%%%%%%%%%%
as well as those of $Q$ and $Q_S$, 
to a given order in QCD perturbation theory, and
comparing them with (\ref{tfq}). 
(The external $b$ quarks are taken to be on-shell.)
This matching procedure 
factorizes the perturbatively calculable short-distance contributions
(Wilson coefficients) from the long-distance dynamics, parametrized by
the non-perturbative matrix elements of local $\Delta B=2$ operators.
We do not use heavy quark effective theory (HQET) to expand these
matrix elements explicitly in $1/m_b$. They are to be understood
in full QCD, in accordance with the treatment of
$1/m_b$ effects in \cite{BBD1}.

In our discussion we shall first concentrate on the contribution to
${\cal T}$ from the operators $Q_1$ and $Q_2$ in (\ref{hpeng}).
The penguin operators have small coefficients and are numerically
less important. Their effect will be included later on. Thus for the time
being only the diagrams $D_1$ -- $D_{10}$ in Fig.~\ref{fignlo} 
are considered, while $D_{11}$ and $D_{12}$ belong to the penguin
sector to be discussed below.

Working in leading order, the matching calculation for
(\ref{tfq}) has to be performed to zeroth order in $\alpha_s$.
At next-to-leading order the coefficients $C_{1,2}$ have to 
be taken  in next-to-leading logarithmic approximation \cite{ACMP,BW} 
and the ${\cal O}(\alpha_s)$ matching corrections to the
coefficients in (\ref{tfq}) have to be computed.
At this order the coefficients in (\ref{tfq}) depend on
the renormalization scheme chosen for their evaluation. This scheme 
dependence is cancelled by the matrix elements of the operators
in (\ref{tfq}), which have to be determined accordingly.

The scheme we employ for our result is specified as follows.
We use dimensional regularization with anti-commuting $\gamma_5$
and modified minimal (${\overline{MS}}$) subtraction of ultraviolet
singularities (NDR scheme).
In addition we project $D$-dimensional Dirac-structures, arising at
intermediate stages of the calculation, according to the prescriptions
($D=4-2\varepsilon$)
\begin{equation}\label{ev1}
[\gamma^\mu\gamma^\alpha\gamma^\nu(1-\gamma_5)]_{ij}\,
[\gamma_\mu\gamma_\alpha\gamma_\nu(1-\gamma_5)]_{kl}
\rightarrow
(16-4\varepsilon) [\gamma^\mu(1-\gamma_5)]_{ij}\,
[\gamma_\mu(1-\gamma_5)]_{kl},
\vspace*{0.1cm}
\end{equation}
\begin{equation}\label{ev2}
[\gamma^\mu\gamma^\alpha\gamma^\nu(1-\gamma_5)]_{ij}\,
[\gamma_\nu\gamma_\alpha\gamma_\mu(1-\gamma_5)]_{kl}
\rightarrow
(4-8\varepsilon) [\gamma^\mu(1-\gamma_5)]_{ij}\,
[\gamma_\mu(1-\gamma_5)]_{kl},
\vspace*{0.2cm}
\end{equation}
\begin{equation}\label{ev3}
[\gamma^\mu\gamma^\nu(1-\gamma_5)]_{ij}\,
[\gamma_\mu\gamma_\nu(1-\gamma_5)]_{kl}
\rightarrow
(8-4\varepsilon) [1-\gamma_5]_{ij}\,[1-\gamma_5]_{kl}
-(8-8\varepsilon) [1-\gamma_5]_{il}\,[1-\gamma_5]_{kj}.
\end{equation}
The projections are equivalent to the subtraction of evanescent
operators, defined by the difference of the left- and right-hand 
sides of (\ref{ev1}--\ref{ev3}). The definition of evanescent
operators is discussed in great detail in \cite{BW,DG,HN2} and we 
use the basis of \cite{HN2} for the projection.
The prescriptions (\ref{ev1}--\ref{ev3}) complete the definition
of our renormalization scheme.
They preserve Fierz symmetry, i.e.
the one-loop matrix elements of $Q$, $Q_S$ and $\tilde Q_S$ equal the 
matrix elements of the operators that one obtains from 
$Q$, $Q_S$, $\tilde Q_S$ by (4-dimensional) Fierz transformations. 
The projections (\ref{ev1}--\ref{ev3}) are sufficient if we use
the Fierz form of $Q_1$, $Q_2$ in (\ref{q1q2}) that corresponds to
closed charm-quark loops in Fig.~\ref{fignlo}. Since the renormalization
of $Q_1$, $Q_2$ respects Fierz symmetry, this choice can always be made.

We now give the result for the transition operator (\ref{tfq})
at next-to-leading order, still neglecting the penguin sector.
The coefficients in (\ref{tfq}) can be written as
\begin{equation}\label{fz}
F(z)=F_{11}(z) C^2_1(\mu_1)+ F_{12}(z) C_1(\mu_1) C_2(\mu_1)+
     F_{22}(z) C^2_2(\mu_1),
\end{equation}
\begin{equation}\label{fij}
F_{ij}(z)=F^{(0)}_{ij}(z)
          +\frac{\alpha_s(\mu_1)}{4\pi}F^{(1)}_{ij}(z)
\end{equation}
and similarly for $F_S(z)$.
The leading order functions $F^{(0)}_{ij}$, $F^{(0)}_{S,ij}$
read explicitly 
\begin{equation}\label{f011}
F^{(0)}_{11}(z)=3\sqrt{1-4z}\, (1-z)\qquad
F^{(0)}_{S,11}(z)=3\sqrt{1-4z}\, (1+2z),
\end{equation}
\begin{equation}\label{f012}
F^{(0)}_{12}(z)=2\sqrt{1-4z}\, (1-z)\qquad
F^{(0)}_{S,12}(z)=2\sqrt{1-4z}\, (1+2z),
\end{equation}
\begin{equation}\label{f022}
\,F^{(0)}_{22}(z)=\frac{1}{2}(1-4z)^{3/2}\qquad\qquad
F^{(0)}_{S,22}(z)=-\sqrt{1-4z}\, (1+2z).
\end{equation}
The next-to-leading order expressions 
$F^{(1)}_{ij}$, $F^{(1)}_{S,ij}$ are
\begin{eqnarray}\label{f111}
F^{(1)}_{11}(z) &=&
32(1-z)(1-2z)\left({\rm Li}_2(\sigma^2)+\ln^2\sigma+
\frac{1}{2}\ln\sigma \ln(1-4z)-\ln\sigma \ln z\right) \nonumber \\
&+& 64(1-z)(1-2z)\left({\rm Li}_2(\sigma)+
\frac{1}{2}\ln(1-\sigma)\ln\sigma\right) \nonumber \\
&-& 4(13-26z-4z^2+14z^3)\ln\sigma \nonumber \\
&+&\sqrt{1-4z}\Bigl[4(13-10z)\ln z-12(3-2z)\ln(1-4z) \nonumber \\
&& +\frac{1}{6}(109-226z+168z^2)\Bigr] 
+ 2\sqrt{1-4z}(5-8z)\ln\frac{\mu_2}{m_b},
\end{eqnarray}
\begin{eqnarray}\label{fs111}
F^{(1)}_{S,11}(z) &=&
32(1-4z^2)\left({\rm Li}_2(\sigma^2)+\ln^2\sigma+
\frac{1}{2}\ln\sigma \ln(1-4z)-\ln\sigma \ln z\right) \nonumber \\
&+& 64(1-4z^2)\left({\rm Li}_2(\sigma)+
\frac{1}{2}\ln(1-\sigma)\ln\sigma\right) \nonumber \\
&-& 16(4-2z-7z^2+14z^3)\ln\sigma \nonumber \\
&+&\sqrt{1-4z}\Bigl[64(1+2z)\ln z-48(1+2z)\ln(1-4z) \nonumber \\
&& -\frac{8}{3}(1-6z)(5+7z)\Bigr] 
- 32\sqrt{1-4z}(1+2z)\ln\frac{\mu_2}{m_b},
\end{eqnarray}
\begin{eqnarray}\label{f112}
F^{(1)}_{12}(z) &=&
\frac{64}{3}(1-z)(1-2z)\left({\rm Li}_2(\sigma^2)+\ln^2\sigma+
\frac{1}{2}\ln\sigma \ln(1-4z)-\ln\sigma \ln z\right) \nonumber \\
&+& \frac{128}{3}(1-z)(1-2z)\left({\rm Li}_2(\sigma)+
\frac{1}{2}\ln(1-\sigma)\ln\sigma\right) \nonumber \\
&+& (2-259z+662z^2-76z^3-200z^4)\frac{\ln\sigma}{6z} \nonumber \\
&-&\sqrt{1-4z}\left[(2-255z+316z^2)\frac{\ln z}{6z}+8(3-2z)\ln(1-4z)
\right. \nonumber \\
&&\left.+\frac{2}{9}(127-199z-75z^2)\right] \nonumber \\
&-& 2\sqrt{1-4z}(17-26z)\ln\frac{\mu_1}{m_b}
+\frac{4}{3}\sqrt{1-4z}(5-8z)\ln\frac{\mu_2}{m_b},
\end{eqnarray}
\begin{eqnarray}\label{fs112}
F^{(1)}_{S,12}(z) &=&
\frac{64}{3}(1-4z^2)\left({\rm Li}_2(\sigma^2)+\ln^2\sigma+
\frac{1}{2}\ln\sigma \ln(1-4z)-\ln\sigma \ln z\right) \nonumber \\
&+& \frac{128}{3}(1-4z^2)\left({\rm Li}_2(\sigma)+
\frac{1}{2}\ln(1-\sigma)\ln\sigma\right) \nonumber \\
&+& (1-35z+4z^2+76z^3-100z^4)\frac{4\ln\sigma}{3z} \nonumber \\
&-&\sqrt{1-4z}\left[(1-33z-76z^2)\frac{4\ln z}{3z}+32(1+2z)\ln(1-4z)
\right. \nonumber \\
&&\left.+\frac{4}{9}(68+49z-150z^2)\right] \nonumber \\
&-& 16\sqrt{1-4z}(1+2z)\ln\frac{\mu_1}{m_b}
-\frac{64}{3}\sqrt{1-4z}(1+2z)\ln\frac{\mu_2}{m_b},
\end{eqnarray}
\begin{eqnarray}\label{f122}
F^{(1)}_{22}(z) &=&
\frac{4}{3}(4-21z+2z^2)\left({\rm Li}_2(\sigma^2)+\ln^2\sigma+
\frac{1}{2}\ln\sigma \ln(1-4z)-\ln\sigma \ln z\right) \nonumber \\
&+& \frac{4}{3}(1-2z)(5-2z)\left({\rm Li}_2(\sigma)+
\frac{1}{2}\ln(1-\sigma)\ln\sigma\right) \nonumber \\
&-& (7+13z-194z^2+304z^3-64z^4)\frac{\ln\sigma}{6z}-
\frac{\pi^2}{3}(1-10z) \nonumber \\
&+&\sqrt{1-4z}\left[(7+27z-250z^2)\frac{\ln z}{6z}-4(1-6z)\ln(1-4z)
\right. \nonumber \\
&&\left.-\frac{1}{18}(115+632z+96z^2)\right] \nonumber \\
&-& 2\sqrt{1-4z}(5-2z)\ln\frac{\mu_1}{m_b}
+\frac{4}{3}\sqrt{1-4z}(2-5z)\ln\frac{\mu_2}{m_b},
\end{eqnarray}
\begin{eqnarray}\label{fs122}
F^{(1)}_{S,22}(z) &=&
-\frac{32}{3}(1+z)(1+2z)\left({\rm Li}_2(\sigma^2)+\ln^2\sigma+
\frac{1}{2}\ln\sigma \ln(1-4z)-\ln\sigma \ln z\right) \nonumber \\
&+& \frac{32}{3}(1-4z^2)\left({\rm Li}_2(\sigma)+
\frac{1}{2}\ln(1-\sigma)\ln\sigma\right) \nonumber \\
&+& (1+7z+10z^2-68z^3+32z^4)\frac{4\ln\sigma}{3z}+
\frac{8\pi^2}{3}(1+2z) \nonumber \\
&-&\sqrt{1-4z}\left[(1+9z+26z^2)\frac{4\ln z}{3z}-16(1+2z)\ln(1-4z)
\right. \nonumber \\
&&\left.+\frac{8}{9}(19+53z+24z^2)\right] \nonumber \\
&-& 16\sqrt{1-4z}(1+2z)\ln\frac{\mu_1}{m_b}
+\frac{32}{3}\sqrt{1-4z}(1+2z)\ln\frac{\mu_2}{m_b}.
\end{eqnarray}
In these equations we have set $N_c=3$ and used
\begin{equation}\label{li2sig}
{\rm Li}_2(x)=-\int^{x}_{0}dt\frac{\ln(1-t)}{t},\quad\quad
\sigma=\frac{1-\sqrt{1-4z}}{1+\sqrt{1-4z}}, \quad\quad 
z=\frac{m_c^2}{m_b^2}.
\end{equation}
The dependence on the renormalization scale $\mu_1$
in (\ref{f111}--\ref{fs122}) cancels against the scale
dependence of the $\Delta B=1$ Wilson coefficients $C_i(\mu_1)$
in (\ref{fz}) to the considered order in $\alpha_s$. Likewise,
the dependence on $\mu_2$ is cancelled by the matrix elements of
the $\Delta B=2$ operators $Q$ and $Q_S$.
To check this, we note that the scale dependence of matrix elements
$\langle\vec Q\rangle=
(\langle Q\rangle, ~\langle Q_S\rangle, ~\langle\tilde Q_S\rangle)^T$
is given by
\begin{equation}\label{qmu2}
\frac{d}{d\ln\mu_2}\langle\vec Q\rangle=-\frac{\alpha_s}{4\pi}
\gamma^{(0)}_2 \langle\vec Q\rangle ,
\end{equation}
where
\begin{equation}\label{ga0}
\gamma^{(0)}_2=
\left(\begin{array}{ccc}
      4 & 0 & 0 \\
 0 & -28/3 & 4/3 \\
 0 &  16/3 & 32/3 
\end{array}\right).
\end{equation}
The operator $\tilde Q_S$, which is redundant at leading power in $1/m_b$,  
can then be eliminated using (\ref{qstas}) below and the $\mu_2$-independence 
can be verified.

Our results in (\ref{f111}-\ref{fs122}) correspond to the use of the 
one-loop pole mass for the $b$ quark in (\ref{tfq}). In $z=m_c^2/m_b^2$
there is no difference between the ratios of pole masses and 
${\overline{MS}}$ masses at next-to-leading order in $\alpha_s$.

We next present the contribution from QCD penguins to the transition
operator ${\cal T}$. This sector has been treated in \cite{BBD1}
in the leading logarithmic approximation. At next-to-leading order
the penguin coefficients $C_3,\ldots, C_6$ have to be computed with
next-to-leading logarithmic accuracy \cite{BJLW,rome} and the diagram
$D_{11}$ in Fig.~\ref{fignlo} must be evaluated. Because the 
coefficients $C_3,\ldots, C_6$ are small ($\sim$ few per cent),
contributions of second order in these coefficients are safely
negligible and it is sufficient to calculate only the interference
of $C_3,\ldots, C_6$ with $C_1$ and $C_2$. A consistent way to
implement this approximation at NLO is to treat $C_3,\ldots, C_6$
as formally of ${\cal O}(\alpha_s)$. The standard NLO formula
(as reviewed in \cite{BBL}) can be used for the penguin coefficients,
except that terms of order $\alpha_s C_3,\ldots, \alpha_s C_6$
have to be dropped. Accordingly, only current-current operators
$Q_1$, $Q_2$ are inserted into the diagrams $D_1$ -- $D_{11}$. The
NLO result thus obtained is manifestly scheme independent and formally
of order ${\cal O}(C_{3,\ldots,6})=$ ${\cal O}(\alpha_s)$.
A further contribution, absent in leading logarithmic approximation,
comes from the chromomagnetic operator $Q_8$ in (\ref{hpeng})
and is shown as diagram $D_{12}$ of Fig.~\ref{fignlo}.
Since this contribution arises first at ${\cal O}(\alpha_s)$, the
lowest order expression is sufficient for $C_8$.
For the NLO penguin contribution we then find
\begin{equation}\label{tpcq}
{\cal T}_p=-\frac{G^2_F m^2_b}{12\pi}(V^*_{cb}V_{cs})^2
\, \left[ P(z) Q+ P_{S}(z) Q_S \right],
\end{equation}
\begin{equation}\label{pz}
P(z)=\sqrt{1-4z}\left((1-z)K_1'(\mu_1)+\frac{1}{2}(1-4z)K_2'(\mu_1)+
  3z K_3'(\mu_1)\right)
+\frac{\alpha_s(\mu_1)}{4\pi} F_p(z) C^2_2(\mu_1),
\end{equation}
\begin{equation}\label{psz}
P_S(z)=\sqrt{1-4z}(1+2z)\left(K_1'(\mu_1)-K_2'(\mu_1)\right)-
\frac{\alpha_s(\mu_1)}{4\pi} 8 F_p(z) C^2_2(\mu_1),
\end{equation}
\begin{eqnarray}\label{fpz}
F_p(z)&=& -\frac{1}{9}\sqrt{1-4z}(1+2z)\cdot \\
&&\cdot\left[ 2\ln\frac{\mu_1}{m_b}
 +\frac{2}{3}+4z-\ln z+\sqrt{1-4z}(1+2z)\ln\sigma
 +\frac{3 C_8(\mu_1)}{C_2(\mu_1)}\right], \nonumber
\end{eqnarray}
where we defined the combinations
$K_1'=2(3 C_1 C_3+C_1 C_4+C_2 C_3)$, $K_2'=2 C_2 C_4$ and 
$K_3'=2(3 C_1 C_5+C_1 C_6+C_2 C_5+ C_2 C_6)$.
The explicit $\mu_1$-dependence in $F_p(z)$ is cancelled
by the $\mu_1$-dependence of $C_3,\ldots, C_6$. Note that for the
penguin sector the scale and scheme dependence of the matrix elements
of $Q$ and $Q_S$ are effects beyond the considered order and
numerically negligible.

Beyond leading order in the $1/m_b$ expansion, several further 
operators contribute to $\Delta\Gamma$, denoted by $R_i$ in 
\cite{BBD1}. Inspection of the factorized matrix elements of these 
operators given in \cite{BBD1} shows that these superficially 
power-suppressed operators contribute at leading power beyond 
tree level. This is due to the fact that the operators are defined 
in QCD (rather than in HQET) and hence the one-loop matrix elements 
contain $m_b$ explicitly. The leading power piece arises only from 
loop momenta of order $m_b$ and can therefore be subtracted 
perturbatively. This subtraction is necessary for a complete 
calculation of the $\alpha_s$ correction at leading power and is 
taken into account in the above result. To be more explicit, consider 
as an example the operator
\begin{equation}\label{r0def}
R_0\equiv Q_S+\tilde Q_S+\frac{1}{2} Q,
\end{equation}
which can be reduced to an explicitly power-suppressed operator using 
Fierz transformations and the equations of motion. (We have used this 
in our calculation to eliminate $\tilde Q_S$.) At order $\alpha_s$, 
we find the leading power contribution
\begin{eqnarray}\label{qstas}
%%%\tilde Q_S &=& -Q_S-\frac{1}{2}Q \\
\langle R_0 \rangle &=& \frac{\alpha_s}{4\pi}
\left[\left((N_c+1)\ln\frac{\mu}{m_b}+\frac{2-N_c}{N_c}\right) 
\langle Q \rangle \right. \\
&&\left.\quad +\left(4(N_c+1)\ln\frac{\mu}{m_b}+2(N_c+1)\right)
\langle Q_S \rangle \right]+ {\cal O} (1/m_b) \nonumber.
\end{eqnarray}
It is crucial that this relation holds independent of the external state, 
so that power counting is again manifest after subtracting the right hand 
side of (\ref{qstas}) from the matrix element of $R_0$. (The procedure 
discussed here bears some similarity with mixing of higher dimension 
operators into lower dimension operators in cut-off or lattice 
regularizations.) These subtractions 
must be kept in mind when a non-perturbative evaluation of the matrix 
elements of the $R_i$ is combined with the present NLO results. In the 
factorization approximation of \cite{BBD1} these subtractions correspond 
to using the pole $b$ quark mass in the expressions for the factorized 
matrix elements. Indeed, in the 
$N_c\to\infty$ limit, we find \cite{BBD1}
\begin{equation}\label{mer0}
\langle\bar B_s|R_0|B_s\rangle=
f^2_{B_s}M^2_{B_s}\left(1-\frac{M^2_{B_s}}{(\bar m_b+\bar m_s)^2}\right)
\simeq f^2_{B_s}M^2_{B_s}\frac{\alpha_s N_c}{4\pi}
\left(6\ln\frac{m_b}{\mu}-4\right),
\end{equation}
up to corrections of order $(N_c\alpha_s)^2$ and $\Lambda_{\rm QCD}/m_b$. 
The $b$-quark mass $\bar m_b$ in the second expression
of (\ref{mer0}) is the ${\overline{MS}}$ mass at the scale 
$\mu$, which corresponds to our renormalization of the scalar 
operators $Q_S$, $\tilde Q_S$. To obtain the third expression 
we used the 1-loop relation between the pole and ${\overline{MS}}$ 
mass in the large-$N_c$ limit and 
the fact that $M_{B_s}-m_{b,pole}={\cal O}(\Lambda_{\rm QCD})$. 
The same result as (\ref{mer0}) is obtained from the large-$N_c$ limit 
of (\ref{qstas}).
(In deriving (\ref{mer0}) we have used the Fierz transform of 
$\tilde Q_S$. For the coincidence of (\ref{mer0}) with (\ref{qstas}) 
it is crucial that the choice in (\ref{ev3}) 
maintains the Fierz symmetry in the one-loop matrix elements entering 
(\ref{r0def}).)

\vspace*{0.4cm}

%\section{Discussion}

\noindent 
{\bf 3. Discussion.} 
The complete expression for $\Delta\Gamma_{B_s}$ with 
short-distance coefficients at NLO in QCD is given by 
\begin{eqnarray}\label{dgg}
&&\left(\frac{\Delta\Gamma}{\Gamma}\right)_{B_s}=
\frac{16\pi^2 B(B_s\to Xe\nu)}{g(z)\tilde\eta_{QCD}}
\frac{f^2_{B_s}M_{B_s}}{m^3_b}|V_{cs}|^2\cdot \\
&&\cdot\left(G(z)\,\frac{8}{3}\,B+
    G_S(z)\,\frac{M^2_{B_s}}{(\bar m_b+\bar m_s)^2}\,\frac{5}{3}
    \,B_S
   +\sqrt{1-4z} ~\delta_{1/m}\right), \nonumber
\end{eqnarray}
where
\begin{equation}\label{ggs}
G(z)=F(z)+P(z)\quad \mbox{and}\quad G_S(z)=-(F_S(z)+P_S(z)).
\end{equation}
We eliminated the total decay rate $\Gamma_{B_s}$
in favour of the semileptonic branching ratio $ B(B_s\to Xe\nu)$, 
as done in \cite{BBD1}.
This cancels the dependence of $(\Delta\Gamma/\Gamma)$ on $V_{cb}$
and introduces the phase space function
\begin{equation}\label{gz}
g(z)=1-8z+8z^3-z^4-12z^2\ln z,
\end{equation}
as well as the QCD correction factor \cite{CM}
\begin{equation}\label{etqcd}
\tilde\eta_{QCD}=1-\frac{2\alpha_s(m_b)}{3\pi}
\left[\left(\pi^2-\frac{31}{4}\right)(1-\sqrt{z})^2+\frac{3}{2}\right].
\end{equation}
The latter is written here in the approximate form of \cite{KM}.
The bag factors $B$ and $B_S$ parametrize the matrix elements
of $Q$ and $Q_S$,
\begin{eqnarray}\label{qb}
\langle\bar B_s|Q|B_s\rangle &=& \frac{8}{3}f^2_{B_s}M^2_{B_s} B, \\
\label{qsbs}
\langle\bar B_s|Q_S|B_s\rangle &=& -\frac{5}{3}f^2_{B_s}M^2_{B_s}
\frac{M^2_{B_s}}{(\bar m_b+\bar m_s)^2} B_S. 
\end{eqnarray}
The masses $\bar m_b\equiv\bar m_b(\mu)$, $\bar m_s$ refer to the
${\overline{MS}}$ definition. The relation of $\bar m_b$ to
the pole mass $m_b$ is
\begin{equation}\label{mmsbar}
\bar m_b(\mu)=m_b \left[1-\frac{\alpha_s}{\pi}
  \left(\ln\frac{\mu^2}{m^2_b}+\frac{4}{3}\right)\right].
\end{equation}
Finally, $\delta_{1/m}$ describes $1/m_b$ corrections. The
explicit expression for $\delta_{1/m}$ in the factorization approximation
can be found in \cite{BBD1}. In the present NLO 
approximation the $b$-quark mass that appears in
$\delta_{1/m}$ is the pole mass as mentioned above.

For the numerical evaluation we use the following input parameters
(central values):
\begin{equation}\label{mbcs}
m_b=4.8~{\rm GeV}\quad
\bar m_b(m_b)=4.4~{\rm GeV}\quad
z=0.085\quad
\bar m_s=0.2~{\rm GeV},
\end{equation}
\begin{equation}\label{mfbsl}
M_{B_s}=5.37~{\rm GeV}\quad
f_{B_s}=0.21~{\rm GeV}\quad
B(B_s\to Xe\nu)=0.104.
\end{equation}
The two-loop expression is used throughout for the
QCD coupling $\alpha_s$ in the form given in \cite{BBL} with
$\Lambda^{(5)}_{\overline{MS}}=0.225~{\rm GeV}$. The NLO coefficients
$F(z)$, $F_S(z)$ in (\ref{tfq}) and $P(z)$, $P_S(z)$ in (\ref{tpcq}) 
are consistently expanded
to first order in $\alpha_s$. We take
$\mu_1=\mu_2=m_b$ as central values for the renormalization scales.
The dependence of $\Delta\Gamma_{B_s}$ 
on $\bar m_s$ is marginal and its dependence on $z$ stems almost 
totally from $g(z)$ in (\ref{gz}).

\begin{table}[t]
\addtolength{\arraycolsep}{0.2cm}
\renewcommand{\arraystretch}{1.3}
$$
\begin{array}{|c||c|c|c|}
\hline
\mu_1 & m_b/2 & m_b & 2 m_b \\ \hline\hline
G_S=-(F_S+P_S)              & 0.743 & 0.937 & 1.018  \\  \hline
G^{(0)}_S=-(F^{(0)}_S+P^{(0)}_S)  & 1.622 & 1.440 & 1.292 \\  \hline
G=F+P                  & 0.023 & 0.030 & 0.036 \\  \hline
G^{(0)}=F^{(0)}+P^{(0)}      & 0.013 & 0.047 & 0.097 \\  \hline\hline

-F_S                  & 0.867 & 1.045 & 1.111  \\  \hline
-F^{(0)}_S            & 1.729 & 1.513 & 1.341 \\  \hline
F                    & 0.042 & 0.045 & 0.049  \\  \hline
F^{(0)}              & 0.030 & 0.057 & 0.103 \\  \hline
\end{array}
$$
\caption{\label{tabwc}
Numerical values of the Wilson coefficients $G$, $G_S$, $F$, $F_S$,
at next-to-leading order 
in the NDR-scheme with evanescent operators subtracted as described
in the text (for $\mu_2=m_b$). Leading order results (superscript `(0)') 
are also shown for comparison.}
\end{table}
The results for the Wilson coefficients are displayed in 
Table \ref{tabwc}.
The contribution of $Q_S$ dominates $\Delta\Gamma_{B_s}$ since
the coefficient of $Q_S$ is numerically much
larger than the one of $Q$. 
These coefficients are independent of the scale $\mu_1$, related to
the $\Delta B=1$ operators, up to terms of next-to-next-to-leading 
order. As can be seen from Table \ref{tabwc},
the residual $\mu_1$ dependence is indeed substantially decreased
for the coefficient $G$ of $Q$. The reduction of scale
dependence is less pronounced for the coefficient $G_S$ of $Q_S$. 
On the other hand, in our renormalization scheme, the central
value at NLO is considerably smaller (by about $30\%$) than the
leading order result. However, due to the scheme dependence
of the Wilson coefficient, it is premature to draw definitive conclusions
on the size of $\Delta\Gamma_{B_s}$ without combining the 
coefficient functions with $B$ and $B_S$, computed in the same scheme.
For quick reference we rewrite (\ref{dgg}) as 
\begin{equation}\label{dgabc}
\left(\frac{\Delta\Gamma}{\Gamma}\right)_{B_s}=
\left(\frac{f_{B_s}}{210~{\rm MeV}}\right)^2
\left[0.006\, B(m_b)+ 0.150\, B_S(m_b) - 0.063\right], 
\end{equation}
where the numbers are obtained with our central parameter set. 

A preliminary lattice evaluation of the relevant bag parameters 
can be found in \cite{GBS}, together with a complete 1-loop 
lattice-to-continuum matching. We can use their result to obtain 
a (conceptually) complete (but numerically preliminary) 
next-to-leading-order result for 
$(\Delta \Gamma/\Gamma)_{B_s}$. In \cite{GBS} a continuum renormalization 
scheme different from ours for the operators $Q_S$ and $\tilde{Q}_S$ 
is chosen. We computed the relation between the two schemes
to ${\cal O}(\alpha_s)$ and found
\begin{eqnarray}\label{bs45}
B_S &=& B^+_4+\frac{\alpha_s}{4\pi}\left(B^+_4+\frac{1}{15}B^+_5\right),\\
\tilde B_S &=& B^+_5+
\frac{\alpha_s}{4\pi}\left(-\frac{35}{6}B^+_4+\frac{1}{2}B^+_5\right).
\end{eqnarray}
Here $\tilde B_S$ is the bag parameter related to $\tilde Q_S$,
defined in analogy with (\ref{qsbs}) with $Q_S\to\tilde Q_S$,
$B_S\to\tilde B_S$, and the numerical factor $(-5/3)$ $\to$ $(+1/3)$.
$B^+_4$ and $B^+_5$ are the bag parameters $B_S$ and $\tilde B_S$,
respectively, but in the scheme of \cite{GBS}.
From the quoted estimates $B^+_4(\mu_0)=0.80\pm 0.01$,
$B^+_5(\mu_0)=0.94\pm 0.01$ \cite{GBS} (where the errors are statistical
only and $\mu_0=2.33~{\rm GeV}$) we infer
$B_S(\mu_0)=0.81$, $\tilde B_S(\mu_0)=0.87$.
Using (\ref{ga0}), and taking the running of $\bar m_b$ 
(\ref{mmsbar}) into account, the bag parameter $B_S$ at the scale 
$m_b$ is given by
\begin{equation}\label{bsbs0}
B_S(m_b)=B_S(\mu_0)+\frac{\alpha_s}{4\pi}\ln\frac{m_b}{\mu_0}
\left(-\frac{20}{3}B_S(\mu_0)+\frac{4}{15}\tilde B_S(\mu_0)\right) = 
0.75.
\end{equation}
We also take $B(m_b)=0.9$ from the compilation \cite{FS}. 
Without the $1/m_b$ corrections, (\ref{dgabc}) then
becomes $(\Delta\Gamma/\Gamma)_{B_s}=0.118~(f_{B_s}/210~{\rm MeV})^2$.
There is a $\pm 15\%$ error from (continuum) perturbation theory, 
obtained from varying $\mu_1$ between $m_b/2$ and $2 m_b$, and a 
negligible $\pm 1\%$ statistical error from the lattice simulation. 
In addition the sizeable $1/m_b$ correction, 
computed in \cite{BBD1}, has to be 
included. The estimate $-0.063$ in (\ref{dgabc}) is obtained using 
factorization of hadronic matrix elements
and has a relative error of at least $\pm 20\%$. 
As a preliminary result we may therefore write
\begin{equation}\label{dgnum}
\left(\frac{\Delta\Gamma}{\Gamma}\right)_{B_s}=
\left(\frac{f_{B_s}}{210~{\rm MeV}}\right)^2
\left(0.054^{+0.016}_{-0.032}~\mbox{($\mu_1$-dep.)}\,\pm\, ??? 
~\mbox{(latt. syst.)}\right).
\end{equation}
We emphasize the preliminary nature of the central value which 
depends crucially on the estimate $B_4^+(\mu_0)=0.80$ taken from 
\cite{GBS}. There is an unspecified systematic error (indicated by 
the question marks) attached to this number, related to the fact 
that the lattice calculation has been performed in the 
quenched approximation, at finite lattice spacing 
and with a ``$b$-quark'' mass in the charm quark mass region, without 
extrapolation to the continuum limit and to realistic $b$ quark masses, 
respectively. Clearly, for further progress improved lattice
determinations of bag parameters, most importantly of $B_S$, 
are mandatory, and the numerical value for $(\Delta \Gamma/\Gamma)_{B_s}$ 
above has to be regarded in this context. 

The rather low number for the central value, compared to the leading order 
analysis of \cite{BBD1}, is a consequence of the fact 
that $1/m_b$ effects and penguin
contributions \cite{BBD1}, as well as NLO QCD corrections, all lead to
a reduction of $\Delta\Gamma_{B_S}$. This is further reinforced
by the small value $B_S(m_b)=0.75$ in our example.

It is interesting to consider the ratio of $\Delta\Gamma_{B_s}$
to the mass difference $\Delta M_{B_s}$ \cite{BP,BBD1}, in which
the dependence on the decay constant $f_{B_s}$ cancels out and the
sensitivity to $V_{cb}$ is considerably reduced. In addition
$(\Delta\Gamma/\Delta M)_{B_s}$ only depends on the ratio of
bag parameters. Generalizing the results given in \cite{BBD1} to
include the next-to-leading order QCD corrections we can write
\begin{eqnarray}\label{dgdm}
&&\left(\frac{\Delta\Gamma}{\Delta M}\right)_{B_s}=\frac{\pi}{2}
\frac{m^2_b}{M^2_W}\left|\frac{V_{cb}V_{cs}}{V_{ts}V_{tb}}\right|^2
\frac{1}{\eta_B S_0(x_t)}\cdot \\
&&\cdot\left(\frac{8}{3}\,G(z)+
    \frac{5}{3}\frac{M^2_{B_s}}{(\bar m_b+\bar m_s)^2}\,G_S(z)
    \,\frac{B_S}{B}
   +\sqrt{1-4z} ~\delta_{1/m}\right), \nonumber
\end{eqnarray}
where $\eta_B$ is the (scheme-dependent) next-to-leading order
QCD factor entering $\Delta M_{B_s}$ \cite{BJW}. In the usual
NDR scheme $\eta_B(m_b)=0.846$. The top-quark mass dependent function
$S_0(x_t)=S_0((\bar m_t/M_W)^2)=2.41$ for $\bar m_t=167\,{\rm GeV}$.
In analogy with (\ref{dgnum}) we then have 
\begin{equation}\label{dgmnum}
\left(\frac{\Delta\Gamma}{\Delta M}\right)_{B_s}=
\left(2.63^{+0.67}_{-1.36}~\mbox{($\mu_1$-dep.)}\,\pm\, ??? 
~\mbox{(latt. syst.)}\right)\cdot 10^{-3}.
\end{equation}

The next-to-leading order calculation
presented in this article can also be applied, by taking the 
limit $z\to 0$, to the mixing-induced CP asymmetry
in inclusive $B_d(\bar B_d)\to\bar uu\bar dd$ decays \cite{BBD2}. 
The time-dependent asymmetry is given by
${\cal A}(t)={\rm Im}\xi\sin\Delta M t$, where the coefficient
${\rm Im}\xi$ is a measure of the CKM parameter $\sin 2\alpha$. With
next-to-leading order QCD corrections included, the expression
for ${\rm Im}\xi$ in Eq. (25) of \cite{BBD2} is modified to read
\begin{equation}\label{imxi}
{\rm Im}\xi=-0.12\sin 2\alpha\!\left(\frac{f_B}{180~{\rm MeV}}\right)^{\!2}
\left[0.14 B+ 0.64 B_S - 0.07 - 0.06\, 
\frac{\sin\alpha\sin(\alpha+\beta)}{\sin\beta\sin 2\alpha}\right]\!.
\end{equation}  
\noindent In this equation the bag factors $B$ and $B_S$ 
(taken at $\mu=m_b$) are 
the $B_d$ analogues
of those defined in (\ref{qb}) and (\ref{qsbs}) for $B_s$. A detailed 
discussion of CP asymmetries will be presented in \cite{BBGLN2}.

\vspace*{0.4cm}

\noindent 
{\bf Acknowledgements.} 
We thank Guido Martinelli for comments on the manuscript.

\vfill\eject


\begin{thebibliography}{99}
\bibitem{LO}
J.S. Hagelin, Nucl. Phys. {\bf B193}, 123 (1981);
E. Franco, M. Lusignoli and A. Pugliese, 
Nucl. Phys. {\bf B194}, 403 (1982);
L.L. Chau, Phys. Rep. {\bf 95}, 1 (1983);
A.J. Buras, W. S\l ominski and H. Steger, 
Nucl. Phys. {\bf B245}, 369 (1984);
M.B. Voloshin, N.G. Uraltsev, V.A. Khoze and M.A. Shifman, 
Sov. J. Nucl. Phys. {\bf 46}, 112 (1987);
A. Datta, E.A. Paschos and U. T\"urke, 
Phys. Lett. {\bf B196}, 382 (1987);
A. Datta, E.A. Paschos and Y.L. Wu,
Nucl. Phys. {\bf B311}, 35 (1988).
\bibitem{BIG}
I. Bigi {\it et al.},
in {\em B Decays}, second edition, ed. S. Stone, p.~132
(World Scientific, Singapore, 1994) [hep-ph/9401298].
\bibitem{DUN1}
I. Dunietz, Phys. Rev. {\bf D52}, 3048 (1995).
\bibitem{DF}
R. Fleischer and I. Dunietz, Phys. Lett. {\bf B387}, 361 (1996);
Phys. Rev. {\bf D55}, 259 (1997).
\bibitem{BP}
T.E. Browder and S. Pakvasa, 
Phys. Rev. {\bf D52}, 3123 (1995).
\bibitem{BBD1}
M. Beneke, G. Buchalla and I. Dunietz, 
Phys. Rev. {\bf D54}, 4419 (1996).
\bibitem{GRO} 
Y. Grossman, Phys. Lett. {\bf B380}, 99 (1996).
\bibitem{BU1}
I. Bigi and N. Uraltsev, Phys. Lett. {\bf B280}, 271 (1992).
\bibitem{BSUV}
I. Bigi, M. Shifman, N. Uraltsev and A. Vainshtein,
[hep-ph/9805241].
\bibitem{ALE93} 
R. Aleksan {\it et al.}, Phys. Lett. {\bf B316}, 567 (1993).
\bibitem{BBGLN2}
M. Beneke, G. Buchalla, C. Greub, A. Lenz and U. Nierste,
in preparation.
\bibitem{PDG}
C. Caso et al., Particle Data Group, Eur. Phys. J. {\bf C3}, 
1 (1998).
\bibitem{BBL}
G. Buchalla, A.J. Buras and M.E. Lautenbacher,
Rev. Mod. Phys. {\bf 68}, 1125 (1996).
\bibitem{ACMP}
G. Altarelli, G. Curci, G. Martinelli and S. Petrarca,
Nucl. Phys. {\bf B187}, 461 (1981).
\bibitem{BW}
A.J. Buras and P.H. Weisz, Nucl. Phys. {\bf B333}, 66 (1990).
\bibitem{DG}
M.J. Dugan and B. Grinstein, Phys. Lett. {\bf B256}, 239 (1991).
\bibitem{HN2}
S. Herrlich and U. Nierste, Nucl. Phys. {\bf B455}, 39 (1995).
\bibitem{BJLW}
A.J. Buras, M. Jamin, M.E. Lautenbacher and P.H. Weisz,
Nucl. Phys. {\bf B370}, 69 (1992); Addendum-ibid. 
{\bf B375}, 501 (1992).
\bibitem{rome}
M. Ciuchini, E. Franco, G. Martinelli, L. Reina, Nucl. Phys. 
{\bf B415}, 403 (1994).
\bibitem{CM}
N. Cabibbo and L. Maiani, Phys. Lett. {\bf B79}, 109 (1978);
Y. Nir, Phys. Lett. {\bf B221}, 184 (1989).  
\bibitem{KM}
C.S. Kim and A.D. Martin, Phys. Lett. {\bf B225}, 186 (1989).
\bibitem{GBS}
R. Gupta, T. Bhattacharya and S.R. Sharpe,
Phys. Rev. {\bf D55}, 4036 (1997). 
\bibitem{FS}
J.M. Flynn and C.T. Sachrajda, [hep-lat/9710057],
to appear in {\it Heavy Flavours (2nd ed.)}, eds. 
A.J. Buras and M. Lindner, World Scientific, Singapore.
\bibitem{BJW}
A.J. Buras, M. Jamin and P.H. Weisz, Nucl. Phys. {\bf B347}, 491 (1990).
\bibitem{BBD2}
M. Beneke, G. Buchalla and I. Dunietz, 
Phys. Lett. {\bf B393}, 132 (1997).
\end{thebibliography}
\end{document}